\documentclass[aps,prl,twocolumn,superscriptaddress]{revtex4-2}

\usepackage{amsmath}
\usepackage{amssymb}
\usepackage{graphicx}
\usepackage{verbatim}
\usepackage{color}

\begin{document}


\title{Bounce Of Nothing}

\author{Patrick Draper}
\email[]{pdraper@illinois.edu}
\affiliation{Department of Physics, University of Illinois, Urbana, IL 61801}

\author{Isabel Garcia Garcia}
\email[]{isabel@kitp.ucsb.edu}
\affiliation{Kavli Institute for Theoretical Physics, University of California, Santa Barbara, CA 93106}

\author{Benjamin Lillard}
\email[]{blillard@illinois.edu}
\affiliation{Department of Physics, University of Illinois, Urbana, IL 61801}

\date{\today}

\begin{abstract}
Theories with compact extra dimensions are sometimes unstable to decay into a bubble of nothing --- an instability resulting in the destruction of spacetime.
We investigate the existence of these bubbles in theories where the moduli fields that set the size of the extra dimensions are stabilized at a positive vacuum energy --- a necessary ingredient of any theory that aspires to describe the real world. Using bottom-up methods, and focusing on a five-dimensional toy model, we show that four-dimensional de Sitter vacua admit bubbles of nothing for a wide class of stabilizing potentials. We show that, unlike ordinary Coleman-De Luccia tunneling, the corresponding decay rate remains non-zero in the limit of vanishing vacuum energy.
Potential implications include a lower bound on the size of compactified dimensions.
\end{abstract}


\maketitle

\section{\label{sec:intro} I.~Introduction}

A wide range of experimental evidence strongly indicates that our Universe is entering a phase of accelerated expansion \cite{Perlmutter:1998np,Percival:2009xn,Aghanim:2018eyx}. Accommodating this observation in the context of general relativity requires the presence of dark energy, a near-perfect fluid with equation of state $p / \rho \approx - 1$. The simplest realization of dark energy is through a de Sitter solution of Einstein's equations, featuring a constant positive vacuum energy.

Notwithstanding this apparent simplicity, the theoretical standing of de Sitter vacua remains contentious. Constructing de Sitter solutions in string theory has proven remarkably hard \cite{Maldacena:2000mw,Townsend:2003qv,Hertzberg:2007wc,Wrase:2010ew,Shiu:2011zt,Bena:2014jaa,Kutasov:2015eba,Andriot:2016xvq,Moritz:2017xto,Sethi:2017phn,Danielsson:2018ztv,Hamada:2018qef,Hamada:2019ack,Carta:2019rhx,Gao:2020xqh}.
Although there exist prescriptions on how to achieve this goal \cite{Kachru:2003aw,Balasubramanian:2005zx}, the hurdles to overcome have even led to speculation that de Sitter vacua may not exist in theories of quantum gravity \cite{Obied:2018sgi,Garg:2018reu,Ooguri:2018wrx}. Less controversial is the expectation that if de Sitter exists, it will only be metastable.
In any geometric compactification of string theory that aspires to describe the real world, all moduli fields associated with the shape and size of the compactification must be stabilized at a positive value of the scalar potential --- a goal that can only be achieved locally. As a result, a four-dimensional de Sitter vacuum generally will  be unstable to decay into a Universe with vanishing energy density, with spatial extra dimensions that are no longer compactified \cite{Dine:1985he,Giddings:2003zw,Giddings:2004vr}.

In a seminal publication, Coleman and De Luccia (CDL) extended previous results on the topic of tunneling in quantum field theory \cite{Coleman:1977py,Callan:1977pt} to include the effects of gravitation \cite{Coleman:1980aw}. The decay rate per unit volume of the false vacuum is of the form $\Gamma \sim v^4 e^{- \Delta S}$, with $v$ some typical energy scale, and $\Delta S$ the Euclidean action of the  so-called ``bounce" solution that interpolates between true and false vacuum. For a de Sitter vacuum with energy density $U_{\rm fv}$, separated by a large potential barrier from the true Minkowski phase, $\Delta S$ is given by~\cite{Coleman:1980aw}
\begin{equation}
	\Delta S \simeq \frac{24 \pi^2 m_{\rm Pl}^4}{U_{\rm fv}} .
\label{eq:DeltaSE_CdL}
\end{equation}
Eq.(\ref{eq:DeltaSE_CdL}) diverges in the limit $U_\text{fv} \rightarrow 0$, making the corresponding false vacuum exponentially long-lived.

In this letter, we discuss an additional instability of de Sitter and Minkowski vacua that may be present in theories with stabilized extra dimensions.
This decay is a generalization of Witten's ``bubble of nothing" \cite{Witten:1981gj} --- a catastrophic instability that destroys the spacetime.
Like CDL, we focus on a real scalar field, $\phi$, minimally coupled to gravity in four dimensions. Interpreting $\phi$ as the radial modulus of an extra dimension compactified on an $S^1$, we show that a wide range of scalar potentials with a local de Sitter vacuum are compatible with the existence of a bubble of nothing. The instanton describing this instability corresponds to a solution to the CDL equations with unusual boundary conditions. We construct this instanton, analytically and numerically, for a general class of potentials. Evaluating the corresponding Euclidean action, we find that, unlike Eq.~(\ref{eq:DeltaSE_CdL}), it remains finite in the limit of vanishing vacuum energy.

Our result suggests the possibility of an upper bound on the lifetime of realistic de Sitter vacua, set by the size of the extra dimensions, and largely insensitive to the size of the four-dimensional cosmological constant.

Most prior work  has focused on the existence of  bubbles of nothing from the top-down in string compactifications \cite{Fabinger:2000jd,Brill:1991qe,DeAlwis:2002kp,Acharya:2019mcu}, and on the role of supersymmetry in suppressing decay rates \cite{Horowitz:2007pr,Blanco-Pillado:2016xvf, GarciaEtxebarria:2020xsr}.
A number of interesting papers have studied  theories with explicit moduli stabilization mechanisms, both with Anti-de Sitter~\cite{Horowitz:2007pr,BlancoPillado:2010df,BlancoPillado:2010et,Brown:2010mf,Ooguri:2017njy} and de Sitter~\cite{BlancoPillado:2010et,Brown:2010mf} false vacua, and with bubble solutions characterized by shrinking $S^1$~\cite{BlancoPillado:2010df,Horowitz:2007pr} or $S^2$~\cite{BlancoPillado:2010et,Brown:2010mf,Ooguri:2017njy} fibers. We take a complementary, bottom-up approach, remaining agnostic about the details of the stabilizing potential, and we build on the insights of \cite{Dine:2004uw}, where it was first suggested that these instabilities may survive moduli stabilization. Our goal is to develop four-dimensional tools to assess the relevance of bubbles of nothing in phenomenologically viable models.

\section{II.~Gravitation and tunneling}

We dedicate this section to reviewing the relevant aspects of the CDL formalism, as well as Witten's bubble.

\subsection{Coleman-DeLuccia formalism}

The formalism of \cite{Coleman:1977py,Callan:1977pt,Coleman:1980aw} is centered upon finding the bounce: the solution to the Euclidean field equations whose analytic continuation provides the spacetime to which the false vacuum decays. In the presence of gravity, the problem is greatly simplified by the (postulated) $O(4)$ symmetry of the bounce, which restricts our attention to metrics of the form
\begin{equation}
	d s_4 = d \xi^2 + \rho (\xi)^2 d \Omega_3 ,
\label{eq:CdL_metric}
\end{equation}
where $\xi$ is a radial coordinate, and $d \Omega_3$ is the line element of the unit three-sphere, with curvature radius $\rho$.
For a single real scalar field minimally coupled to gravity, the Euclidean field equations take the form
\begin{gather}
	\label{eq:CdL_phi}	\phi'' + \frac{3 \rho'}{\rho} \phi' = \frac{d U (\phi)}{d \phi} , \\
	\label{eq:CdL_rho}	\rho'^2 = 1 + \frac{\rho^2}{6 m_{\rm Pl}^2} \left( \phi'^2 - 2 U (\phi) \right) .
\end{gather}

What are the boundary conditions that accompany Eq.~(\ref{eq:CdL_phi}) and (\ref{eq:CdL_rho})?
The answer depends on the topology of the solution, which as usual in general relativity we don't necessarily know beforehand.
For a de Sitter false vacuum, and a scalar potential satisfying $U(\phi) \geq 0$, the solution has the topology of a four-sphere, and $\rho$ vanishes twice \cite{Guth:1982pn}.
WLOG, the first zero can be placed at $\xi = 0$, and $\xi \in [0, \xi_{\rm max}]$, with $\rho (\xi_{\rm max}) \equiv  0$. Any non-singular solution must then satisfy the boundary conditions $ \phi'(0) = \phi'(\xi_{\rm max}) = 0$.

One such solution is the de Sitter false vacuum:
\begin{equation}
	\phi_{\rm dS} \equiv \phi_{\rm fv} , \qquad {\rm and} \qquad \rho_{\rm dS}(\xi) = \Lambda \sin \left( \frac{\xi}{\Lambda} \right) ,
\label{eq:dS}
\end{equation}
with $\Lambda \equiv \sqrt{3 m_{\rm Pl}^2 / U_{\rm fv}}$ the radius of the cosmological de Sitter horizon. On the other hand, the bounce describing the decay of this false vacuum features an inner region where $\phi$ approaches the true vacuum, and an outside region where the solution asymptotes to the metastable phase. The exponential factor governing the decay rate is obtained by evaluating the action of the bounce relative to that of the false vacuum, that is $\Delta S \equiv \left. S_E \right|_{\rm CDL} - \left. S_E \right|_{\rm dS}$, and  the action evaluated on any non-singular solution to the field equations can be written as $S_E = - 2 \pi^2 \int_0^{\xi_{\rm max}} d \xi \rho^3 U$. When the de Sitter vacuum energy is much smaller than the potential barrier separating false and true vacua, $\Delta S$ is given by Eq.~(\ref{eq:DeltaSE_CdL}).

\subsection{Bubble of nothing}

Witten's bubble is a peculiar instability of the Kaluza-Klein vacuum $\mathbb{M}^4 \times S^1$ \cite{Witten:1981gj}.
The corresponding instanton is the five-dimensional Euclidean Schwarzschild solution,
\begin{equation}
	d s_5 =  \frac{dr^2}{1 - R^2/r^2 } + r^2 d \Omega_3 + \left( 1 - \frac{R^2}{r^2} \right) dy^2 ,
\label{eq:5DES}
\end{equation}
with $r \in [R, \infty)$, and $y \sim y + 2 \pi R$.
Analytically continuing Eq.~(\ref{eq:5DES}) into a Lorentzian manifold reveals the spacetime into which the Kaluza-Klein vacuum decays. The new spacetime resembles the original $\mathbb{M}^4 \times S^1$, except with a `hole' inside: from the perspective of a four-dimensional observer, the region ${\bf x}^2 - t^2 < R^2$ has been removed, and spacetime ends on the surface ${\bf x}^2 - t^2 = R^2$. The existence of a fifth dimension plays a crucial role in ensuring that the solution remains smooth. Although asymptotically the proper length of the $S^1$ remains $2 \pi R$, it shrinks to zero size at ${\bf x}^2 - t^2 = R^2$ --- effectively sealing off what otherwise would be a manifold with a boundary. Like in the traditional picture of quantum tunneling, a bubble is nucleated within the false vacuum sea. Unlike the familiar process, the bubble is not filled with true vacuum, but rather it has `nothing' in it. 

It was noted in \cite{Dine:2004uw} that Witten's bubble can be rewritten as a solution of Eq.~(\ref{eq:CdL_phi}) and (\ref{eq:CdL_rho}). Indeed, the five-dimensional Einstein-Hilbert action can be dimensionally reduced into a four-dimensional problem including Einstein's gravity plus a real scalar $\phi$ --- the radial modulus, whose vacuum expectation value determines the size of the compactified dimension. The following parametrization results in the canonical normalization for $\phi$:
\begin{equation}
	d s_5 	= e^{-\sqrt{\frac{2}{3}} \frac{\phi}{m_{\rm Pl}}} d s_4 + e^{2 \sqrt{\frac{2}{3}} \frac{\phi}{m_{\rm Pl}}} dy^2 .
\label{eq:5D_red}
\end{equation}
Witten's bubble can be written as in Eq.~(\ref{eq:5D_red}), with the four-dimensional metric satisfying the $O(4)$-symmetric ansatz of Eq.~(\ref{eq:CdL_metric}). The corresponding field equations are Eq.~(\ref{eq:CdL_phi}) and (\ref{eq:CdL_rho}), with $U (\phi) \equiv 0$.

\begin{figure}[h]
\includegraphics{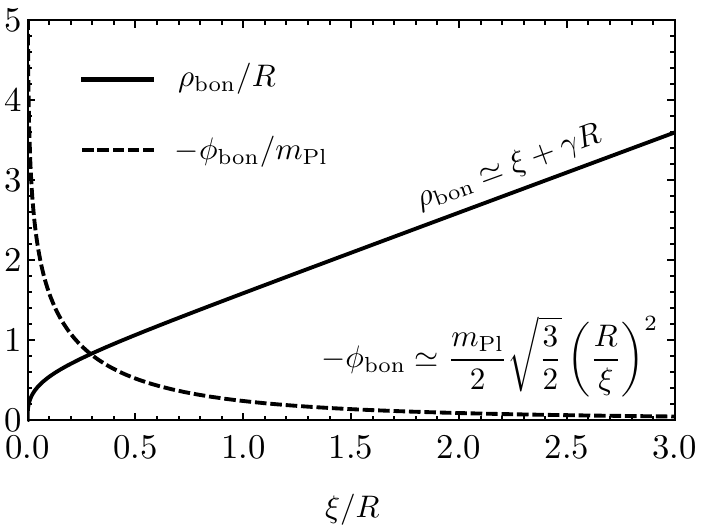}
\caption{\label{fig:BON} Witten's bubble, in terms of the degrees of freedom of the corresponding CDL problem. Its asymptotic behavior, for $\xi \gg R$, is shown in the figure, with $\gamma = \sqrt{\pi} \Gamma(\frac{3}{4}) /\Gamma(\frac{1}{4}) \simeq 0.6$.}
\end{figure}

The $O(4)$-invariant CDL coordinate is related to the radial coordinate of the Schwarzschild metric by
\begin{equation}
	\xi (r) = \int_R^r \frac{d \hat r}{ (1 - R^2 / {\hat r}^2)^{1/4} } ,
\end{equation}
and the profiles $\phi (\xi)$ and $\rho (\xi)$ corresponding to Witten's bubble are shown in Fig.~\ref{fig:BON}.
Near $\xi = 0$:
\begin{gather}
	\label{eq:BON_phi_xi_small} \phi_{\rm bon} (\xi) \simeq m_{\rm Pl} \sqrt{\frac{2}{3}} \log \left( \frac{3\xi}{2R} \right) , \\
	\label{eq:BON_rho_xi_small} \rho_{\rm bon} (\xi) \simeq R \left( \frac{3\xi}{2R} \right)^{1/3} .
\end{gather}
This highlights how Witten's bubble is not a CDL bounce proper, as it appears singular at $\xi = 0$. The existence of a fifth dimension is key to ensure that this is only a coordinate singularity. In the near-horizon region:
\begin{equation}
	d s_5 \simeq d \lambda^2 + \lambda^2 d \tilde y^2 + R^2 d \Omega_3 ,
\label{eq:near_horizon}
\end{equation}
where $\lambda \equiv R \left( 3 \xi / 2 R \right)^{2/3}$, and $\tilde y \equiv y / R$. The solution is indeed smooth, and has the topology of $\mathbb{R}^2 \times S^3$.

Upon dimensional reduction, the action of Witten's bubble consists of a single term evaluated at $\xi = 0$:
\begin{equation} \begin{split}
	\Delta S_{\rm bon} 	& = \pi^2 m_{\rm Pl} \sqrt{\frac{2}{3}} \rho_{\rm bon} (\xi)^3 \phi'_{\rm bon}  (\xi) \Big|_{\xi = 0} \\
				& = \pi^2 m_{\rm Pl}^2 R^2 .
\label{eq:S_BON}
\end{split} \end{equation}

\section{III.~Bounce of nothing}

It was an important insight of \cite{Dine:2004uw} that rewriting the bubble of nothing as a CDL problem makes it natural to look for analogous solutions in the presence of a non-vanishing $U(\phi)$. Indeed, in any realistic construction, all moduli must be stabilized, so the survival of these solutions in the presence of a scalar potential is a necessary requirement if they are to remain relevant in `real life'.

What classes of scalar potentials are compatible with the existence of Witten's bubble? It was speculated in \cite{Dine:2004uw} that only potentials that vanish in the compactification limit would be compatible with a bubble of nothing. Instead, let us consider potentials whose asymptotic behavior is of the form $U(\phi) \simeq U_0 \exp \{a \phi / m_{\rm Pl} \}$ as $\phi \rightarrow - \infty$, with no assumption regarding the sign of $a$. Evaluated on Witten's solution, the right-hand-side of Eq.~(\ref{eq:CdL_phi}) near $\xi = 0$ reads
\begin{equation}
	\left. \frac{d U (\phi)}{d \phi} \right|_{\rm bon} \simeq \frac{a U_0}{m_{\rm Pl}} \left( \frac{3 \xi}{2 R} \right)^{\sqrt{\frac{2}{3}} a} .
\end{equation}
On the other hand, both terms on the left-hand side of Eq.~(\ref{eq:CdL_phi}) scale as $\xi^{-2}$ in the limit $\xi \rightarrow 0$. Solutions satisfying the same boundary conditions as Witten's bubble at the center of the bounce are therefore compatible with the corresponding field equations, provided
\footnote{The same conclusion follows from inspecting Eq.~(\ref{eq:CdL_rho}).}
\begin{equation}
	\sqrt{\frac{2}{3}} a + 2 > 0 \quad \Rightarrow \quad a > - \sqrt{6} .
\label{eq:gamma_bound}
\end{equation}
This includes all $a > 0$, in which case the potential vanishes as $\xi \rightarrow 0$, where the extra dimension shrinks to zero size. More surprisingly, it also allows for a negative range of $a$, corresponding to $|U| \rightarrow \infty$ in the direction where the extra dimension becomes compactified. 

\begin{figure}
\includegraphics{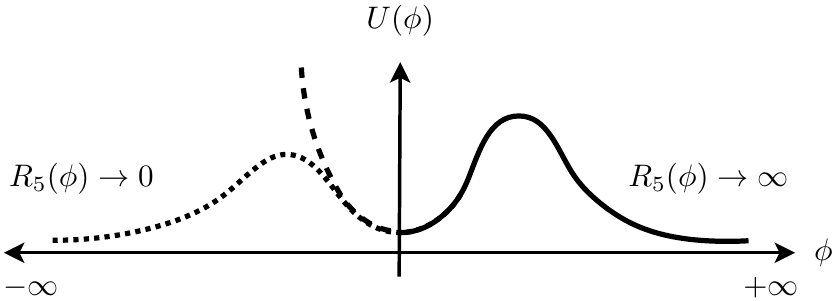}
\caption{\label{fig:U} Theories with stabilized extra dimensions and positive vacuum energy are generally unstable to decay into the decompactification regime, as illustrated by the $\phi \geq 0$ region of this figure. Instead, our work concerns the region $\phi \leq 0$ that probes the compactification limit. Depending on the details of the underlying theory, the asymptotic behavior of the potential in the compactification regime may vary, as illustrated by the dashed and dotted curves.}
\end{figure}

In the remainder of this letter we focus on potentials that satisfy the requirement of Eq.~(\ref{eq:gamma_bound}). As illustrated in Fig.~\ref{fig:U}, this includes potentials that vanish in the compactification limit, but also potentials that grow exponentially provided the growth rate is not too large.
For simplicity, we will take the mass of the modulus in the false vacuum to satisfy $m \ll 1/R$, and assume that the overall scale of the potential is such that $U_0 \lesssim m_{\rm Pl}^2 m^2 \ll m_{\rm Pl}^2 / R^2$.
These simplifications are not necessary for the existence of these instantons, or any of their qualitative features we discuss, but it will allow us to obtain simple analytic expressions for both the bounce and the corresponding tunneling exponent. A more general discussion, where these assumptions are relaxed, can be found in \cite{dSdecays}.
The location of the false vacuum is arbitrary, and we choose $\phi_{\rm fv} \equiv 0$. Around the metastable vacuum, $U (\phi) \simeq U_{\rm fv} + \frac{1}{2} m^2 \phi^2$.

There is a one-parameter family of bubble of nothing solutions to Eq.~(\ref{eq:CdL_phi}) and (\ref{eq:CdL_rho}) with $U(\phi) \equiv 0$, given by
\begin{gather}
	\label{eq:phi_eta} \phi_\eta (\xi) \equiv \sqrt{\frac{3}{2}} m_{\rm Pl} \log \eta + \phi_{\rm bon} (\xi \cdot \eta^{-3/2}) , \\
	\label{eq:rho_eta} \rho_\eta (\xi) \equiv \eta^{3/2} \rho_{\rm bon} (\xi \cdot \eta^{-3/2}) ,
\end{gather}
for any $\eta > 0$. The value of $\eta$, however, is without consequence, as it merely rescales the five-dimensional metric by a constant conformal factor. Given that Euclidean Schwarzschild is a solution of Einstein's equations in vacuum, one may therefore take $\eta \equiv 1$ WLOG. A more extensive discussion on this point can be found in section~1 of the \emph{Supplemental Material}.

The above no longer holds in the presence of a potential for $\phi$. In this case, Eq.~(\ref{eq:phi_eta}) and (\ref{eq:rho_eta}) only provide an approximate solution to the Euclidean field equations for a limited range of $\xi \geq 0$. Nevertheless, as long as $y \sim y + 2 \pi R$, the corresponding five-dimensional geometry remains smooth. Indeed, near $\xi = 0$,
\begin{equation}
	ds_5 \simeq d \lambda^2 + \lambda^2 d \tilde y^2 + \eta^2 R^2 d \Omega_3 ,
\label{eq:near_horizon_eta}
\end{equation}
with $\lambda$ and $\tilde y$ as defined below Eq.~(\ref{eq:near_horizon}). Eq.~(\ref{eq:near_horizon_eta}) clarifies the meaning of $\eta$: it parametrizes the radius of the $S^3$ of the near-horizon geometry, $R_3 \equiv \eta R$, corresponding to the radius of the `hole' that nucleates in spacetime.

Provided $U_0 \ll m_{\rm Pl}^2/R^2$, it is easy to see by inspecting Eq.~(\ref{eq:CdL_phi}) and (\ref{eq:CdL_rho}) that Witten's bubble provides an approximate solution well into the regime where $\rho \gtrsim R$. Here, $\rho' \simeq 1$, and
\begin{equation}
	\phi_\eta (\rho) \simeq \sqrt{\frac{3}{2}} m_{\rm Pl} \left\{ \log \eta - \frac{\eta^3}{2} \left( \frac{R}{\rho} \right)^2 \right\} .
\end{equation}
Assuming that $\eta - 1 \ll 1$, as we will justify in time, $| \phi | / m_{\rm Pl} \ll 1$, and the system finds itself in the vicinity of the false vacuum.
In this regime, Eq.~(\ref{eq:CdL_phi}) reads
\begin{equation}
	\frac{d^2 \phi}{d \rho^2} + \frac{3}{\rho} \frac{d \phi}{d \rho} \simeq m^2 \phi ,
\end{equation}
and the relevant solution is of the form
\begin{equation}
	\phi_i (\rho) \simeq C \frac{K_1 (m \rho)}{m \rho} ,
\label{eq:phi_i}
\end{equation}
where $C$ is a constant of integration.
When $m \rho \ll 1$, $| \phi_i (\rho) | \propto \rho^{-2}$, whereas
\begin{equation}
	| \phi_i (\rho) | \propto \frac{e^{-m \rho}}{(m \rho)^{3/2}} \qquad {\rm for} \qquad m \rho \gg 1 .
\end{equation}
$\phi_i$ interpolates between the bubble of nothing and the metastable vacuum, approaching the latter exponentially fast once $\rho \gg m^{-1}$.
Demanding that both $\phi$ and $\phi'$ remain continuous across the transition allows us to find expressions for both $C$ and $\eta$. In particular,
\begin{equation}
	\eta = \frac{R_3}{R} \simeq 1 + \frac{m^2 R^2}{4} \log (m R)^{-1} + \mathcal{O} \left( m^2 R^2 \right) .
\label{eq:eta}
\end{equation}
By assumption, $m R \ll 1$, which justifies our earlier approximation that $\eta - 1 \ll 1$. The terms of $\mathcal{O} \left( m^2 R^2 \right)$ capture the dependence of the solution on the detailed features of the potential in the region to the left of the local minimum, and, as it will become apparent in section~IV, they do not affect the leading contribution to the bounce action.
Further details complementing this discussion can be found in section~1 of the \emph{Supplemental Material}.

Provided $U_{\rm fv} > 0$, the bounce eventually transitions into the de Sitter false vacuum, where $\rho$ is given by
\begin{equation}
	\rho(\xi) \simeq \Lambda \sin \left( \frac{\xi + \gamma R}{\Lambda} \right) ,
\end{equation}
with $\gamma = \sqrt{\pi} \Gamma(\frac{3}{4}) /\Gamma(\frac{1}{4}) \simeq 0.6$.

\begin{figure}
	\includegraphics{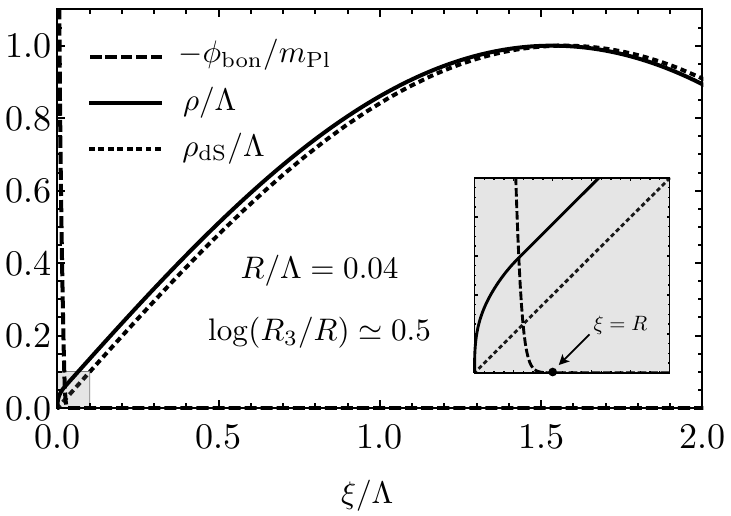}
	\caption{\label{fig:BON_FV} Instanton describing a bubble of nothing instability of a de Sitter vacuum. The choice of scalar potential used in this analysis is given in section~2 of the \emph{Supplemental Material}. The radius of the extra dimension in the false vacuum, $R$, is an input, as indicated. The radius of the corresponding bubble of nothing, $R_3$, is obtained via an overshoot-undershoot algorithm.}
\end{figure}

Our conclusions  are borne out by numerical analysis. Like the traditional CDL bounce, this class of instantons can be found by implementing an overshoot-undershoot method, where $\phi$ corresponds to the position of a particle moving in a potential $-U(\phi)$. In the familiar CDL solution, $\phi$ starts at rest in the vicinity of the true vacuum, corresponding to the boundary condition $\phi'(0) = 0$, and eventually approaches the local minimum. On the contrary, the bubble of nothing starts at $\phi (0) = - \infty$, with ``infinite velocity". The particle breezes through the potential, and finally comes to rest in the vicinity of the false vacuum. Unlike the CDL bounce, the appropriate shooting parameter is not $\phi (0)$, but rather $\eta = R_3 / R$.
Fig.~\ref{fig:BON_FV} shows one such numerical solution.
A more extensive numerical discussion can be found in \cite{dSdecays}.

\section{IV.~de Sitter decay rate}

Having built an approximate solution that interpolates between Witten's bubble and the de Sitter false vacuum, we now turn to the evaluation of the corresponding Euclidean action. Since we are interested in the behavior of the action in the limit of vanishing cosmological constant, we will neglect terms that vanish as $U_{\rm fv} \rightarrow 0$. Relative to that of the false vacuum, the bounce action is given by
\begin{equation} \begin{split}
	\Delta S = & \pi^2 m_{\rm Pl} \sqrt{\frac{2}{3}} \rho (\xi)^3 \phi' (\xi) \Big|_{\xi = 0} \\
			& - 2 \pi^2 \int_0^{\xi_{\rm max}} d \xi \rho^3 U - S_E \Big|_{\rm dS} .
\label{eq:SE_full}
\end{split} \end{equation}

The first line in Eq.~(\ref{eq:SE_full}) is a contribution from the center of the bounce. It reads
\begin{equation} \begin{split}
	\Delta S \Big|_{\xi = 0} \		& \equiv \pi^2 m_{\rm Pl} \sqrt{\frac{2}{3}} \rho (\xi)^3 \phi' (\xi) \Big|_{\xi = 0}  \\
							& = \pi^2 m_{\rm Pl}^2 R^2 \eta^3 \\
							& \simeq \pi^2 m_{\rm Pl}^2 R^2 \left\{ 1 + \frac{3}{4} m^2 R^2 \log (m R)^{-1} \right\} ,
\label{eq:DeltaS_1}
\end{split} \end{equation}
where we have used Eq.~(\ref{eq:eta}) to expand around $\eta = 1$, and we have ignored terms of $\mathcal{O} (m^2 R^2)$ inside the bracket.

Let us now turn to the second line in Eq.~(\ref{eq:SE_full}). Much like for the traditional CDL solution, the contribution to the action from the region where the bounce has effectively reached the false vacuum is vanishingly small. We are therefore left with that from the inner region, up to $\rho \sim m^{-1}$. In this regime, the integral is dominated by the region of large $\rho$, where $\rho' \simeq 1$, and $U(\phi)$ is in the vicinity of the local minimum. We find
\begin{equation} \begin{split}
	\Delta S \Big|_U \	& \equiv - 2 \pi^2 \int_0^{\xi_{\rm max}} d \xi \rho^3 U - S_E \Big|_{\rm dS}  \\
					& \simeq - 2 \pi^2 \int^{m^{-1}} d \rho \rho^3 ( U - U_{\rm fv}) \\
					& \simeq - \frac{3 \pi^2}{8} m_{\rm Pl}^2 m^2 R^4 \log (m R)^{-1},
\label{eq:DeltaS_2}
\end{split} \end{equation}
where, again, we ignore non-log-enhanced terms, as well as terms that vanish as $U_{\rm fv} \rightarrow 0$.

Adding up Eq.~(\ref{eq:DeltaS_1}) and (\ref{eq:DeltaS_2}), the bounce action in the limit $U_{\rm fv} \rightarrow 0$ is approximated by
\begin{equation}
	\Delta S \simeq \pi^2 m_{\rm Pl}^2 R^2 \left\{ 1 + \frac{3}{8} m^2 R^2 \log (m R)^{-1} \right\} .
\label{eq:DeltaS_final}
\end{equation}
This result is finite, and smaller than that of the traditional CDL bounce, given in Eq.~(\ref{eq:DeltaSE_CdL}).
Under the simplifying assumptions introduced in section~III --- namely, that the mass of the radion and the overall scale of the potential satisfy $m \ll 1/R$ and $U_0 \lesssim m_{\rm Pl}^2 m^2$ --- the tunneling exponent in Eq.~(\ref{eq:DeltaS_final}) is approximately equal to that of the original bubble of nothing, up to small corrections.
Relaxing these assumptions will change the exact expression for $\Delta S$, but, crucially, the tunneling exponent will remain finite provided the scalar potential satisfies the existence condition discussed around Eq.~(\ref{eq:gamma_bound}). Further details on this point are dicussed in section~3 of the \emph{Supplemental Material}. A more extended discussion of the behavior of the decay rate into a bubble of nothing that relaxes our previous assumptions on the form of the scalar potential can be found in  \cite{dSdecays}. 

\section{V.~Conclusions}

It remains crucial to improve our understanding of de Sitter vacua, and this includes identifying its potential instabilities.
This can affect the viability of proposed de Sitter constructions, with potential implications for our understanding of the string landscape and the cosmological constant problem.

We have shown that de Sitter vacua can be susceptible to a bubble of nothing instability, with a decay rate that remains finite in the limit of vanishing vacuum energy.
This is true for a wide range of stabilizing potentials, including potentials that diverge in the compactification limit, and that naively would only be susceptible to the spontaneous decompactification instability of \cite{Dine:1985he,Giddings:2003zw,Giddings:2004vr}. This opens the possibility that Witten's bubble may be a crucial ingredient to our understanding of vacuum decay in generic and realistic models, and it is the main qualitative result of this work.

There is a lower bound on the action of the corresponding instanton, approximated by that of Witten's bubble, and that is largely independent of the four-dimensional cosmological constant, or the size of the barrier separating false and true vacua. To illustrate the potential implications of this result, let us estimate the decay probability of our Universe into one of these bubbles. Parametrically, it is given by  $\Gamma / H_0^4 \sim v^4 e^{-\Delta S} / H_0^4 $, and $\Delta S \gtrsim 560 - \log ( m_{\rm Pl}^4 / v^4)$ is necessary to ensure that $\Gamma / H_0^4 \lesssim 1$. For a bounce action similar to that of Eq.~(\ref{eq:DeltaS_final}), this leads to a moderate lower bound on the size of the compact dimension: $R \gtrsim 8 m_{\rm Pl}^{-1}$.

Our work has followed a bottom-up philosophy, investigating the existence of bubbles of nothing in phenomenologically relevant settings, namely assuming the stabilization of moduli at a positive vacuum energy.
This is not a sufficient condition for our Universe to admit this class of instabilities, but it is a necessary one.

Obstacles to the existence of bubbles of nothing have mainly concerned supersymmetry, as a choice of supersymmetric boundary conditions for fermions can provide a topological obstruction \cite{Witten:1981gj}, and in cases without such an obstruction the rate must still vanish in the supersymmetric limit~\cite{Horowitz:2007pr,Blanco-Pillado:2016xvf}. However, the topological obstruction is not universal across all internal manifolds, and moreover, supersymmetry is broken in our universe by an amount much greater than required by the positive vacuum energy. Further, non-supersymmetric boundary conditions that generate tree-level fermion masses provide a phenomenologically attractive implementation of supersymmetry breaking \cite{Antoniadis:1998sd,Delgado:1998qr,Pomarol:1998sd,Barbieri:2002uk,Barbieri:2003kn,Dimopoulos:2014aua,Garcia:2015sfa,Abel:2015oxa,Abel:2017vos}, and are compatible with such bubbles \cite{Fabinger:2000jd}. Overall, the obstructions to the existence of bubbles of nothing \emph{in our world} seem less rather than more.


\begin{acknowledgments}
	{\bf Acknowledgments} We thank N.~Craig, M.~Dine, P.~Fox, A.~Hebecker, and M.~Reece for comments and discussions.
	PD and BL acknowledge support from the US Department of Energy under Grant No.~DE-SC0015655.
	The research of IGG is funded by the Gordon and Betty Moore Foundation through Grant GBMF7392, and by the NSF under Grant No.~NSF PHY-1748958.
\end{acknowledgments}

\bibliography{bounce_of_nothing_refs}

\clearpage
\newpage
\maketitle
\onecolumngrid
\begin{center}
\textbf{\large Bounce Of Nothing} \\ 
\vspace{0.05in}
{ \it \large Supplemental Material}\\ 
\vspace{0.05in}
{Patrick Draper, Isabel Garcia Garcia, and Benjamin Lillard}
{ }

\end{center}

\pagenumbering{roman}

\subsection{1.~Generalized bubble of nothing}

The one-parameter family of bubbles of nothing given in Eq.~(\ref{eq:phi_eta}) and (\ref{eq:rho_eta}) correspond to a five-dimensional metric of the form
\begin{equation} \begin{split}
	d s_5 	& = e^{-\sqrt{\frac{2}{3}} \phi_\eta (\xi) m_{\rm Pl}^{-1}} \left( d \xi^2 + \rho_\eta (\xi)^2 d \Omega_3 \right) + e^{2 \sqrt{\frac{2}{3}} \phi_\eta (\xi)m_{\rm Pl}^{-1}} dy^2 \\
			& = \eta^2 \left\{ e^{-\sqrt{\frac{2}{3}} \phi_{\rm bon} (\hat \xi)m_{\rm Pl}^{-1}} \left( d \hat \xi^2 + \rho_{\rm bon} (\hat \xi)^2 d \Omega_3 \right) + e^{2 \sqrt{\frac{2}{3}} \phi_{\rm bon} (\hat \xi)m_{\rm Pl}^{-1}} dy^2 \right\} ,
\end{split} \end{equation}
where $\hat \xi \equiv \xi \cdot \eta^{-3/2}$. The corresponding geometry is identical to that of the original bubble of nothing introduced in section II, except for an overall conformal factor.
In vacuum, its only consequence is a rescaling of the five-dimensional gravitational constant, $G_5 \rightarrow \eta^{-1} G_5$, and one may therefore choose $\eta \equiv 1$ WLOG.

As discussed in the text, $\eta$ acquires physical significance in the presence of a scalar potential. Near $\xi = 0$,
\begin{equation}
	\phi_\eta (\xi) \simeq m_{\rm Pl} \sqrt{\frac{2}{3}} \log \left( \frac{3 \xi}{2 R} \right) ,
	\qquad {\rm and} \qquad
	\rho_\eta (\xi) \simeq \eta R \left( \frac{3 \xi}{2 R} \right)^{1/3} ,
	\label{eq:BON_eta_small}
\end{equation}
whereas for $\xi \gg R$,
\begin{equation}
	\phi_\eta (\xi) \simeq \sqrt{\frac{3}{2}} m_{\rm Pl} \left\{ \log \eta - \frac{\eta^3}{2} \left( \frac{R}{\xi} \right)^2 \right\} ,
	\qquad {\rm and} \qquad
	\rho_\eta (\xi) \simeq \xi + \gamma \eta^{3/2} R .
\end{equation}
Substituting Eq.~(\ref{eq:BON_eta_small}) into Eq.~(\ref{eq:5D_red}) reveals that the corresponding near-horizon geometry is indeed as in Eq.~(\ref{eq:near_horizon_eta}).

Demanding that $\phi$ and $\phi'$ remain continuous as the solution transitions from bubble of nothing to the intermediate behavior of Eq.~(\ref{eq:phi_i}) allows us to obtain the values of $C$ and $\eta$. Imposing continuity at $\rho = \bar \rho \gg R$, we find
\begin{equation}
	C \simeq - m_{\rm Pl} \sqrt{\frac{3}{2}} \frac{\eta^3 R^2}{\bar \rho^2} \frac{1}{K_2 (m \bar \rho)} , \qquad{\rm and}
\end{equation}
\begin{equation}
	\log \eta \simeq \eta^3 \frac{m^2 R^2}{4} \log \left( \frac{2 e^{-\gamma_E}}{m \bar \rho} \right) \simeq \frac{m^2 R^2}{4} \left\{ \log ( m R )^{-1} + \mathcal{O} (1) \right\} .
\end{equation}
In particular, the leading contribution to $\log \eta$ is independent of  the choice of $\bar \rho$, whose exact value will depend on the features of the potential to the left of the false vacuum. \\

\subsection{2.~Toy model}

The example shown in Fig.~\ref{fig:BON_FV} has been obtained by numerically solving Eq.~(\ref{eq:CdL_phi}) and Eq.~(\ref{eq:CdL_rho}). The scalar potential used in our analysis is given by
\begin{equation}
	U(\phi) = U_0 \beta_0 \left( \frac{e^{3 a \phi m_{\rm Pl}^{-1}}}{3} - \frac{1 + \beta_1}{2} e^{2 a \phi m_{\rm Pl}^{-1}} + \beta_1 e^{a \phi m_{\rm Pl}^{-1}} \right) \qquad {\rm for} \qquad \phi \leq 0.
\end{equation}
$\beta_0$ and $\beta_1$ are defined by demanding that the potential features a local minimum at $\phi_{\rm fv} = 0$, and a local maximum at $\phi = \phi_{\rm max}$ such that $U(\phi_{\rm max}) = U_0$. As an expansion in $\delta \equiv U_{\rm fv} / U_0$, $\beta_0 = 81/4 + \mathcal{O} (\delta)$ and $\beta_1 = 1/3 + 8 \delta / 81 + \mathcal{O} (\delta^2)$, whereas the potential barrier is located at $a \phi_{\rm max} / m_{\rm Pl} = - \log 3 + \mathcal{O} (\delta)$. The choice of parameters corresponding to the solution in Fig.~\ref{fig:BON_FV} has been $a = 5$, $\delta = 0.005$, and $R / \Lambda= 0.04$. \\

\subsection{3.~Exponentially growing potentials}

The reader may be concerned that our results may be qualitatively altered if the scalar potential grows in the compactification limit. This is not so. Integrating Eq.~(\ref{eq:CdL_phi}) allows us to estimate the change in $\eta$ arising from the region near the center of the bounce. Expanding around $\eta = 1$, it reads
\begin{equation}
	\delta \eta \approx - \frac{\sqrt{2/3}}{3 M_p R^2} \int_0^R d\xi \, \rho^3 \frac{d U}{d \phi} .
\end{equation}
Using $U (\phi) \simeq U_0 \exp\{a \phi / m_{\rm Pl}\}$ as the behavior of the scalar potential in the compactification limit, and ignoring irrelevant $\mathcal{O}(1)$ factors, we find
\begin{equation}
	\delta \eta \sim - \frac{a}{1 + a / \sqrt{6}} \frac{U_0 R^2}{m_{\rm Pl}^2} .
\end{equation}
$\delta \eta$ grows large in the limit $a \rightarrow - \sqrt{6}$, in which case our approximations are indeed no longer applicable. However, so long as $a$ is not too close to this limiting value, and that $U_0 \ll m_{\rm Pl}^2 / R^2$, $\delta \eta$ will remain small. Moreover, provided $U_0 \lesssim m_{\rm Pl}^2 m^2$ (in keeping with our assumptions, as introduced in section III), $\eta$ is well approximated by Eq.~(\ref{eq:eta}), and the leading contributions to the bounce action are as in Eq.~(\ref{eq:DeltaS_final}). A more elaborated discussion can be found in \cite{dSdecays}.

\end{document}